\documentclass[12pt]{article} 
\usepackage{amsmath,amsfonts,amssymb,graphicx,array,hyperref}
\usepackage{hypcap,a4wide,ae,tabularx,makeidx,color,psfrag,multirow,colordvi,graphicx} 
\usepackage{wrapfig,float,latexsym,textcomp,slashed,epsfig}

\begin{document}


\thispagestyle{empty}
\begin{center}
{\Large \textbf{Single Particle Closed Orbits in Yukawa Potential}}\\
\vspace{1cm}
{\large \textsc{Rupak Mukherjee$^{\dagger \ref{ipr} \ref{rkmvu} \ref{rkmrc}}$, Sobhan Sounda$^{\ddagger \ref{rkmrc}}$}}
\begin{enumerate}
\item \label{ipr} Institute for Plasma Research, HBNI, Gandhinagar, Gujarat, India.
\item \label{rkmvu} Ramakrishna Mission Vivekananda University, Belur Math, Howrah, Kolkata, India.
\item \label{rkmrc} Ramakrishna Mission Residential College (Autonomous), Narendrapur, Kolkata, India.
\end{enumerate}
\end{center}

$\dagger $\texttt{ rupakmukherjee01@gmail.com}\\
$~~~~~~\ddagger $\texttt{ sounda6@gmail.com}
\vspace{1cm}

\today
\vspace{1cm}

{\large \bf {Abstract:}}\\

We study the orbit of a single particle moving under the Yukawa potential and observe the precessing ellipse type orbits. The amount of precession can be tuned through the coupling parameter $\alpha$. With a suitable choice of the coupling parameter; we can get a closed bound orbit. In some cases we have observed some petals which can also have a closed bound nature with an appropriate choice of the coupling constant. A threshold energy has also been calculated for the boundness of the orbits. \\



\newpage

\newpage
\section  {Introduction:}
It is widely known that there are only two types of central potential $\frac{1}{r}$ and $r^2$ in which all finite motions take place in closed paths$[\ref{land}]$. Some exceptions of the statement above have been reported recently$[\ref{prev1},\ref{prev2}]$.\\
On the other hand Yukawa potential or screened coulomb potential of the form $V(r) = -\frac{\alpha}{r}e^{-\frac{r}{\lambda}}$, ( $\lambda$ is the screening parameter that determines the range of this interaction) has a long standing legacy to represent various physical systems. Its application ranges from astronomy, high-energy physics, nuclear physics, condensed matter physics to plasma physics and many other branches of physics. In the domain of plasma physics this potential is used mostly in Strongly Coupled Plasmas which has several applications in modelling the cores of white dwarfs, planetary rings, atmospheric lightning, molten salts and plasma technology. Hence quite naturally one may ask what is the orbit of a single particle under this screened coulomb potential. Can we expect some bound orbits for the non-zero value of $\alpha$?\\
This study is a more general case of the previous studies $[\ref{prev1},\ref{prev2}]$ that has been done till now regarding the closed bound orbits distinct from coulomb or harmonic oscillator potentials because by the truncation of the exponential series in the Yukawa potential one gets back the potentials assumed in the literature so far. Hence this study is a superset of the previous studies performed so far to the best of our knowledge.\\    
The paper is organised as follows. Section 2 discusses the theory of central force motion for any arbitrary potential. Following the theory we have calculated the trajectories for Yukawa potential in section 3 and the thrust has been given on the calculation for threshold energy for bound orbit in Yukawa potential in section 4. In section 5, some possible applications of this study has been mentioned.

\section{A Brief Discussion of Central Force Motion}

The total energy of a particle moving under a central force is given by,\\
\begin{equation*}
\frac{1}{2}m \dot{r}^2 + \frac{J^2}{2mr^2} + V(r) = E, ~~ J = mr^2 \dot{\theta}
\end{equation*}
where m is the mass of the particle, r is the radial distance of the particle from the force center, J is the angular momentum of the particle, V is the potential of the particle and $\theta$ is the angular coordinate of the particle with respect to some reference axis.
For $r = \frac{1}{u}$ i.e. $\dot{r} = -\frac{J}{m} \frac{du}{d\theta}$; the above equation looks like
\begin{equation*}
\frac{1}{2}m \frac{J^2}{m^2}\left(\frac{du}{d\theta}\right)^2 + \frac{J^2u^2}{2m} + V(u) = E
\end{equation*}
For energy to be constant, i.e. $\frac{dE}{d\theta} = 0$,
\begin{eqnarray*}
\frac{d^2u}{d\theta^2} = - \frac{m}{J^2} \frac{d}{du}\left(V + \frac{J^2 u^2}{2m}\right) = -\frac{m}{J^2} \frac{dV_{eff}}{du}
\end{eqnarray*}
where $\frac{dJ}{d\theta} = 0$ since, for central force the angular period $\Delta \theta$ of the radial oscillators $r(\theta)$ is independent of the angular momentum $J$. In dimensionless coordinates, $ u = \frac{u}{a}$ and $ V = \frac{m}{J^2 a^2}V$ the equation of motion becomes,
\begin{equation*}
\frac{d^2u}{d\theta^2} = - \frac{d V_{eff}}{du}
\end{equation*}
with $V_{eff} = V(u) + \frac{1}{2} u^2$. Now if the form of the potential is known, we can find the trajectory of the particle. The threshold energy value for a bound orbit in any central force can be calculated from the total energy ($h$) conservation relation,
\begin{eqnarray}
&& \frac{1}{2}my^2 + V_{eff}(u) = h, ~~ where, ~ V_{eff}(u) = -\int \limits_0^u F(u)du ~~ and ~ y = \frac{du}{d \theta}\nonumber \\
&& \label{ref} \Rightarrow y = \pm \sqrt{\frac{2}{m}[h - V_{eff}(u)]}
\end{eqnarray}
Now, if we differentiate the above equation we get,
\begin{equation}
y \frac{dy}{du} = \frac{1}{m} \frac{dV_{eff}}{du}
\end{equation}
Here, for $\frac{dV_{eff}}{du} = 0$, if $\frac{dy}{du}$ is not undefined, i.e. the curve plotted from equation $(\ref{ref})$, does not cut the $u$ axis perpendicularly, then $y = 0$ which in turn represents $h = V_{eff}$ i.e. the minimum threshold value of energy for bound orbit. \\
Now if $\frac{dy}{du}$ is undefined, i.e. the curve plotted from equation $\ref{ref}$, cuts the $u$ axis perpendicularly, then from the roots $(u_c,0)$ or the critical points of the equation $\frac{dV_{eff}}{du} = 0$ we find some equilibrium points of the system. Now from the relations $\frac{dV(u_c)}{du}= 0$ and $\frac{dV_{eff}}{du} = 0$ we can conclude that $V(u)$ has either a relative extremum or a horizontal inflection point at $u=u_c$.\\
If,\\ 
I. the $V_{eff}(u)$ has a relative minimum at $u=u_c$, then $V_{eff}(u_c) = h_0$ (threshold or minimum value for energy) and the critical point is a centre and is stable.\\
II. the $V_{eff}(u)$ has a relative maximum at $u=u_c$, then the critical point is a saddle point and is unstable.\\
III. the $V_{eff}(u)$ has a horizontal inflection point at $u=u_c$, then the critical point is of a degenerate type called a cusp and is unstable.

\section{Motion of a Single Particle under Yukawa Potential}

The Yukawa or screened coulomb potential is given by:
\begin{eqnarray}
&& V(r) = -\frac{\alpha}{r}e^{-\frac{r}{\lambda}}\\
\Rightarrow && V_{eff}(u) = \frac{1}{2}u^2 - \frac{m\alpha}{J^2a}u e^{-\frac{1}{u}}
\end{eqnarray}
where, $\lambda$ is the range of the Yukawa potential.\\
The equation of motion is given by
\begin{eqnarray}
\label{eom} \frac{d^2u}{d\theta^2} = -u + \alpha e^{-\frac{1}{u}}\left(1+\frac{1}{u}\right), ~~~ where,~~ \alpha = \frac{m\alpha}{J^2a}
\end{eqnarray}
which can be rewritten in terms of energy as:
\begin{eqnarray}
&& \frac{1}{2}my^2 + \frac{1}{2}u^2 - \alpha ue^{-\frac{1}{u}} = h \nonumber \\
\label{en}\Rightarrow && y = \pm \sqrt{2h - u^2 + 2 \alpha ue^{-\frac{1}{u}}}
\end{eqnarray}
where we have chosen $m=1$. The exact values of $\alpha$ differ for different physical cases depending on the nature of application$[\ref{Harbasic},\ref{appln},\ref{solar},\ref{radar}]$. \\

The Equation $(\ref{eom})$ is difficult to solve for an exact analytical solution.  For planetary type motion the equation of motion with our parameters becomes,
\begin{eqnarray}
\label{planatary}\frac{d^2u}{d\theta^2} +u = 1 + \alpha e^{-\frac{1}{\lambda u}}\left(1+\frac{1}{\lambda u}\right)
\end{eqnarray}
In ref$[\ref{analytical}]$ an approximate analytical solution has been obtained by expanding the R.H.S. of $(\ref{planatary})$ in a Taylor series and truncating it to the second order. 
Thus the equation looks like,
\begin{eqnarray*}
&&  \frac{d^2u}{d\theta^2} +u \left[ 1 - \alpha \left(\frac{a_{p0}^2}{\lambda^2}\right)e^{-a_{p0}/\lambda}\right] = u_p\\
&& u_p = \frac{1}{p_0}\left[ 1+\alpha e^{-a_{p0}/\lambda} \left(1+\frac{a_{p0}}{\lambda} - \frac{a_{p0}^2}{\lambda^2} \right) \right]
\end{eqnarray*}
where $a_{p0} \sim p_0 = 1/u_0 $ and $u_0$ is the unperturbed ($\alpha = 0$) solution. 
The solution of the above equation given by,
\begin{eqnarray*}
&& u(\theta) = u_p + u_e \cos \omega(\theta-\theta_0)\\
&& \omega = \left[1-\alpha(a_{p0}^2/\lambda^2)e^{-a_{p0}/\lambda}\right]^{1/2} 
\end{eqnarray*}
is only valid for $\alpha << 1$ and can deviate significantly for particle orbits which lie away from the unperturbed ($\alpha = 0$) value. The aim of our present study is to explore the full solution space of equation $(\ref{eom})$ by retaining the complete nonlinear form of the force term. We obtain these solutions by a numeical solution of the equation.\\


We adopt another analytical tool viz. a linear stability analysis for a particle under yukawa type potential. We proceed to some extent and then turn to the numerical analysis for different $\alpha$ (coupling constant).

\section{Linear Stability Analysis$[\ref{strogatz}]$ for Single Particle Motion in Yukawa Potential}

For the sake of analysis, we add a small viscous term proportional to velocity ($\mu \frac{du}{d\theta}$), in the equation of motion and eventually put $\mu = 0$ for the final calculation.\\
Thus for Yukawa Potential the modified equation of motion $(\ref{eom})$ can be written as,
\begin{eqnarray*}
&& u^{\prime \prime}(\theta) = f(u^\prime(\theta),u(\theta)) = -\mu u^\prime -u + \alpha \left(1+\frac{1}{u}\right)e^{-\frac{1}{u}}\\
\Rightarrow && u^\prime (\theta) = F(u(\theta),y(\theta)) = y\\
&& y^\prime (\theta) = G(u(\theta),y(\theta)) = -\mu y -u + \alpha \left(1+\frac{1}{u}\right)e^{-\frac{1}{u}}
\end{eqnarray*}
Hence, the fixed points $f(y^*,u^*)$ $=$ $0$ are:\\
$(y^*,u^*)$ $=$ $(0,-0.2114)$ for $\alpha = 0.0005$.\\
$(y^*,u^*)$ $=$ $(0,-0.3046)$ for $\alpha = 0.005$.\\
$(y^*,u^*)$ $=$ $(0,-0.4705)$ for $\alpha = 0.05$.\\
$(y^*,u^*)$ $=$ $(0,-0.7296)$ for $\alpha = 0.5$.\\
$(y^*,u^*)$ $=$ $(0,-0.8094)$ for $\alpha = 1$.\\
The Jacobian for the above set of equation will be given by$[\ref{linear}]$,
\begin{eqnarray*}
&& J(u^*,y^*) = \begin{pmatrix}
\frac{\partial F}{\partial u} & \frac{\partial F}{\partial y} \\
\frac{\partial G}{\partial u} & \frac{\partial G}{\partial y}
\end{pmatrix}_{u=u^*,y=y^*}
= \begin{pmatrix} 0 & 1 \\ -1 + \frac{\alpha}{u^3}e^{-\frac{1}{u}} & -\mu \end{pmatrix}_{u=u^*,y=y^*}
\end{eqnarray*}
The Eigenvalues of the above matrix will be,
\begin{eqnarray*}
\lambda_{1,2} = \frac{1}{2} \left(\tau \pm \sqrt{\tau^2-4\Delta} \right) = \frac{1}{2} \left(-\mu \pm \sqrt{\mu^2 - 4\left(1 - \frac{\alpha}{u^3}e^{-\frac{1}{u}}\right)_{u=u^*}}\right)
\end{eqnarray*}
\begin{wrapfigure}{r}{0.5\textwidth}
\includegraphics[scale=0.6]{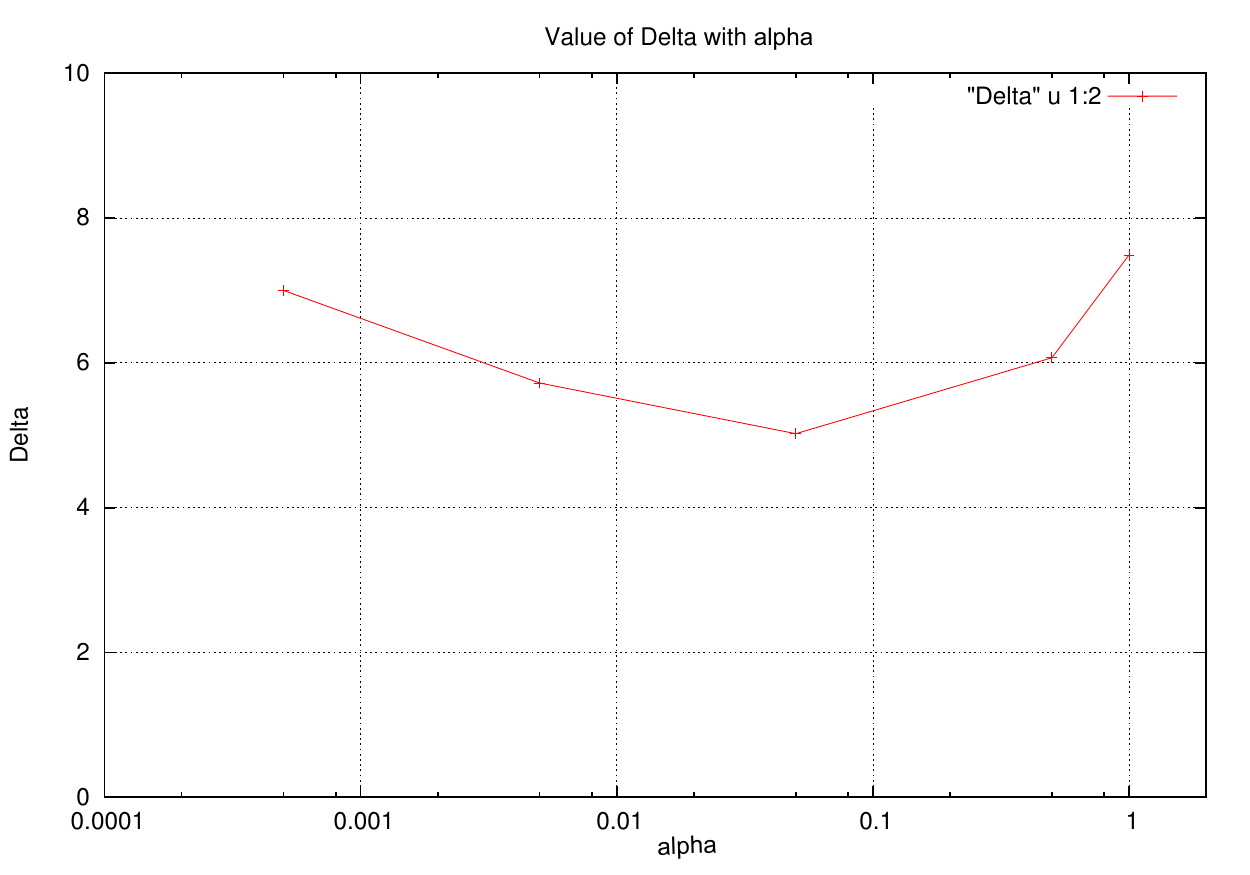}
\end{wrapfigure}
It is checked that the values of $\Delta$ are positive for all values of $\alpha$ given above. Hence, for the above parameter values the fixed point can not be a saddle point.\\
Now, for $\Delta > 0$\\
if $\tau < 0$ and $\tau^2 - 4 \Delta > 0$ then $f(y^*,u^*)$ is a stable node.\\
if $\tau < 0$ and $\tau^2 - 4 \Delta < 0$ then $f(y^*,u^*)$ is a stable spiral.\\
if $\tau > 0$ and $\tau^2 - 4 \Delta > 0$ then $f(y^*,u^*)$ is an unstable node.\\
if $\tau > 0$ and $\tau^2 - 4 \Delta > 0$ then $f(y^*,u^*)$ is an unstable spiral.\\
if $\tau = 0$ and $\tau^2 - 4 \Delta > 0$ then $f(y^*,u^*)$ is a nutrally stable center.\\
The existance of the limit cycle around each of the fixed points (for different values of $\alpha$) has been checked and it is found that for none of the cases there exists any limit cycle.\\
Hence if we continuously change $\mu$ from positive to negative the fixed point changes from stable to unstable spiral. However at $\mu = 0$ we do not have a true hopf bifurcation because there are no limit cycles on either side of the bifurcation. This situation is identical to the case of a damped pendulum or a duffing oscillator.\\ 
Further we analyse the case of $\mu = 0$ numerically for the quantitative understanding of the parameter values with closed orbits.. 

\section{Numerical Analysis for Single Particle Motion in Yukawa Potential}

For numerical analysis, we have used a simple Runge Kutta 4th Order solver and have used standard accepted algorithm for avoiding the numerical singularities, if encountered. Interestingly we have shown that for some typical values of $\alpha$ we get a closed bound orbit. In some cases the orbits have some petals and the number of petals can also be controlled through the judicious choice of the coupling constant. Here we present some results of our simulation that helps to understand the dependency on the coupling constant easily.\\

For $\alpha$ = 0.05; we observe a constant precesion of the ellipse. But the ellipse does not close itself when it completes one revolution. We have tuned the value of the coupling constant to 0.0501 and have observed the expected closure.

\begin{figure}[H]
\includegraphics[scale=0.6]{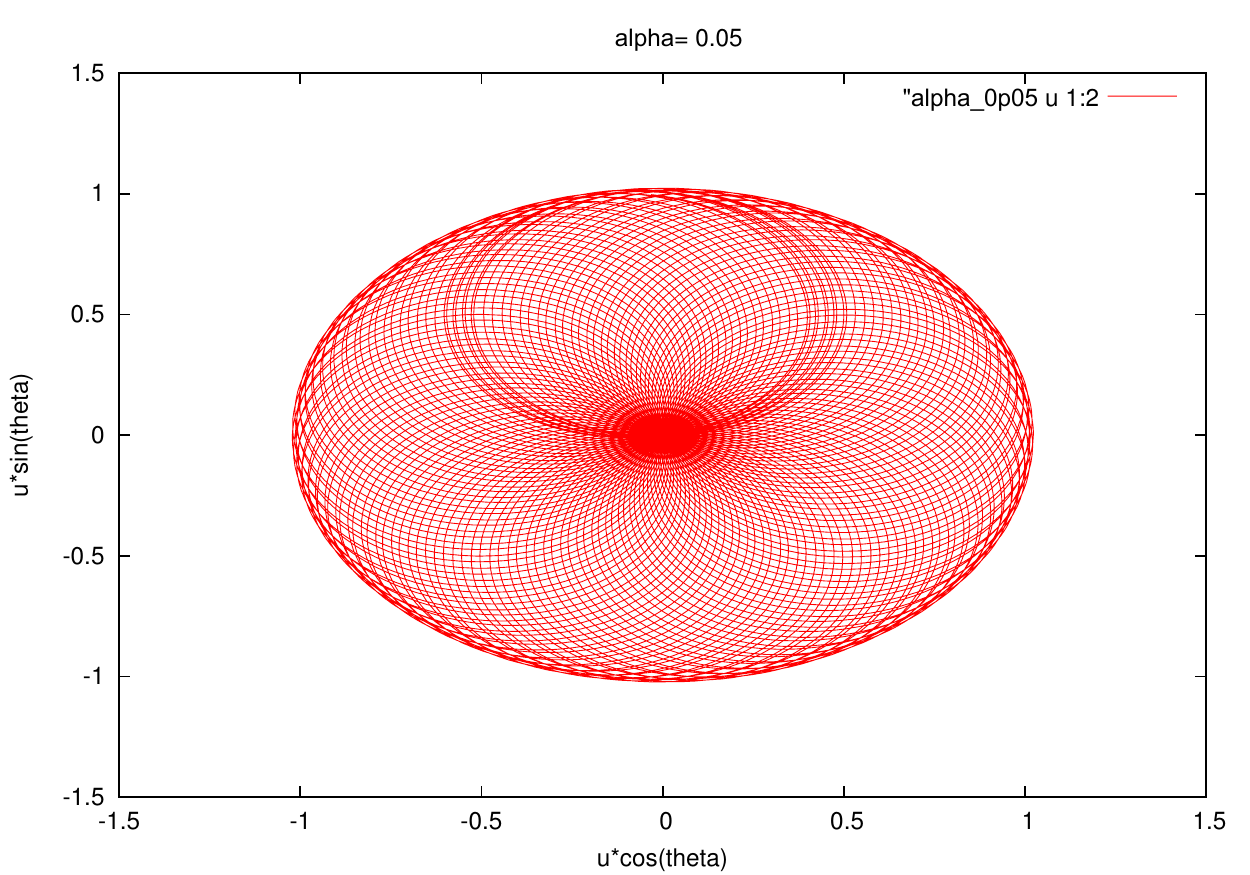}
\includegraphics[scale=0.6]{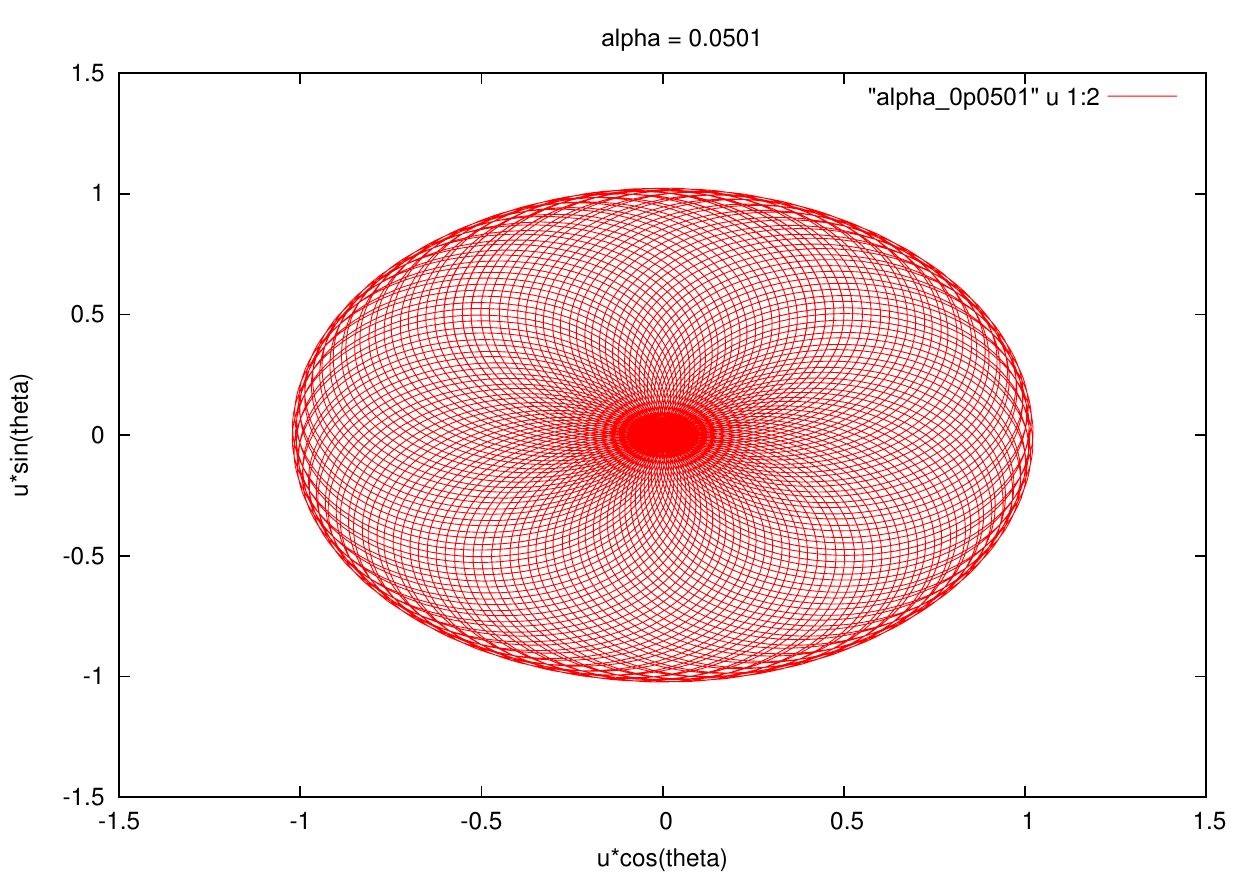}
\caption{Precessing Ellipse with $\alpha=0.05$ and Closed orbits with $\alpha=0.0501$}
\end{figure}

For $\alpha$ = 0.1; we observe almost same behavior but the amount of precesion has increased. Still we had to tune the coupling constant to 0.11 to get a closed bound orbit.

\begin{figure}[H]
\includegraphics[scale=0.6]{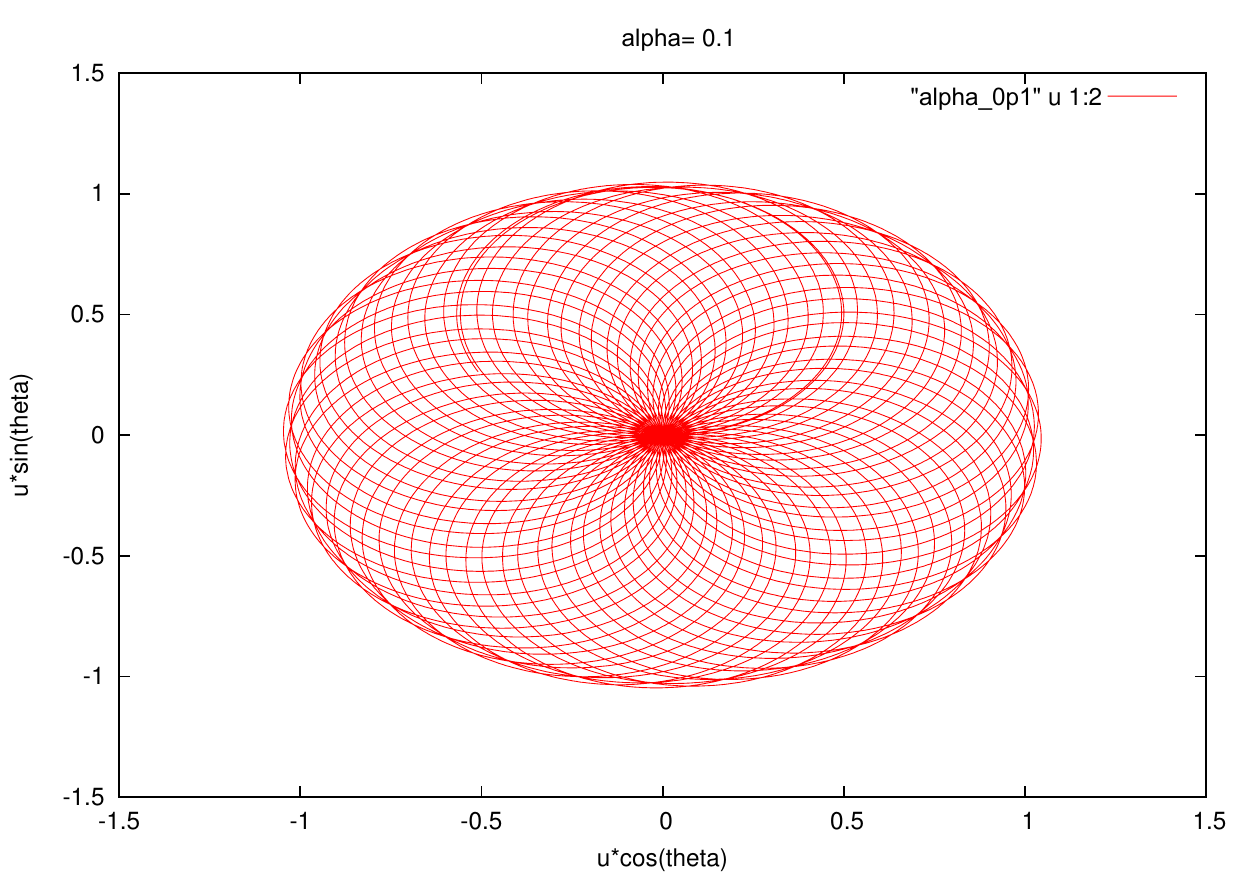}
\includegraphics[scale=0.6]{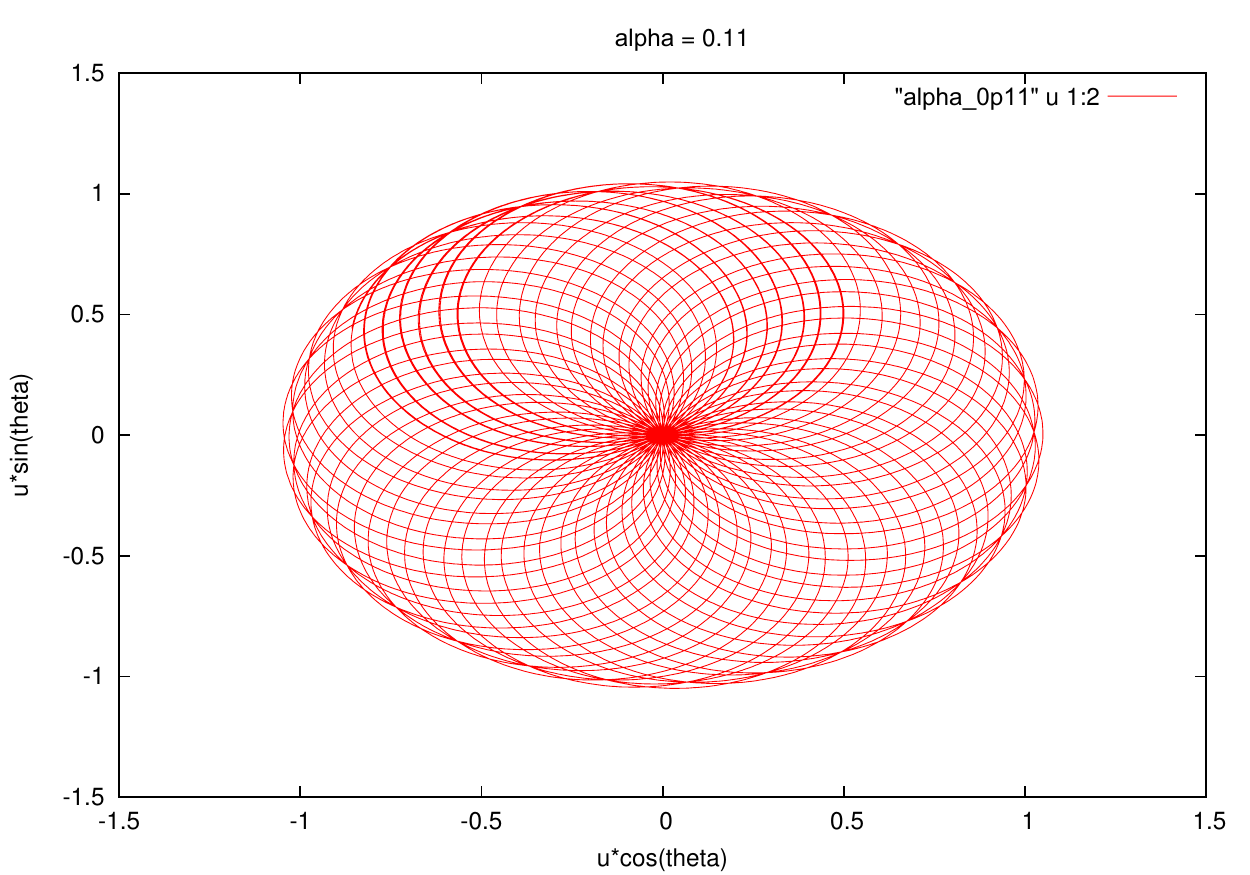} 
\caption{Precessing Ellipse with $\alpha=0.1$ and Closed orbits with $\alpha=0.11$}
\end{figure}

For $\alpha$ = 0.5; we observe a small grouping of the precessed orbits. The precession occurs in a bunch of 2 orbits. Hence we may observe that there are two different values of precession within one complete revolution. When we set the coupling constant to 0.49 we suddenly observe a closed bound orbit maintaining the nature of grouping.

\begin{figure}[H]
\includegraphics[scale=0.6]{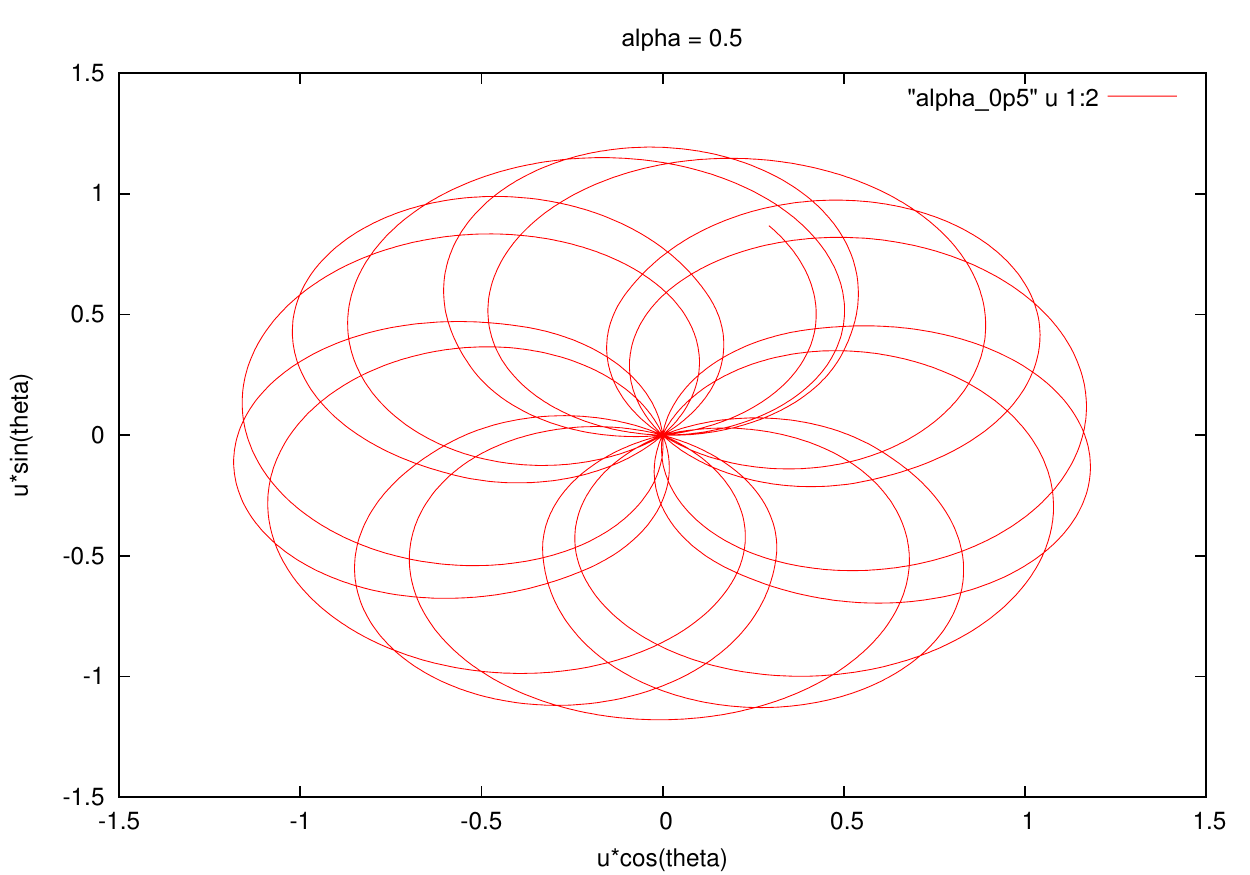}
\includegraphics[scale=0.6]{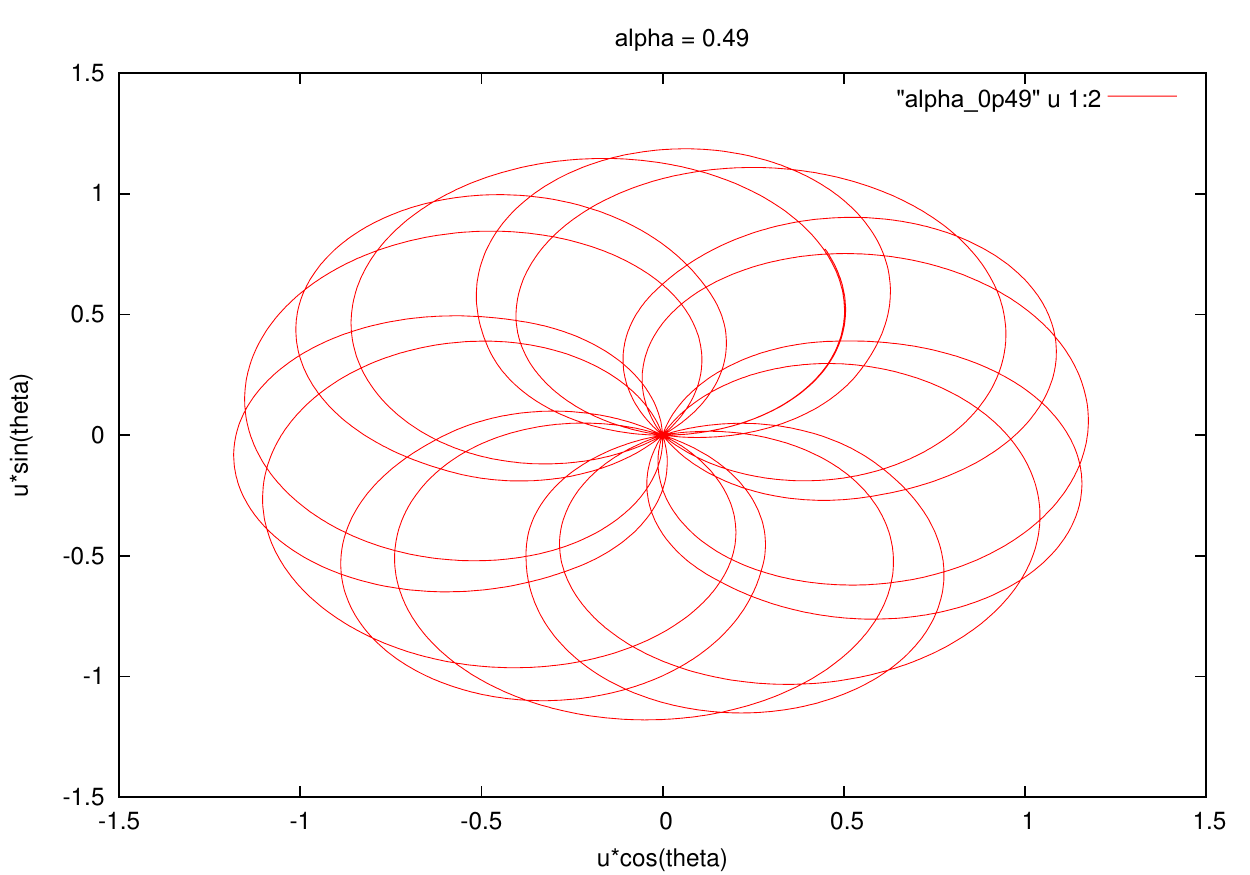}
\caption{Precessing Ellipse with $\alpha=0.5$ and Closed orbits with $\alpha=0.049$}
\end{figure}

For $\alpha$ = 1; we observe two distinct types of orbits, which has been further investigated in the next step. Here a 3-fold larger orbit as well as a 3-fold smaller orbit are found. The tuning of the coupling constant has been done accordingly to obtain the closed bound orbit.

\begin{figure}[H]
\includegraphics[scale=0.6]{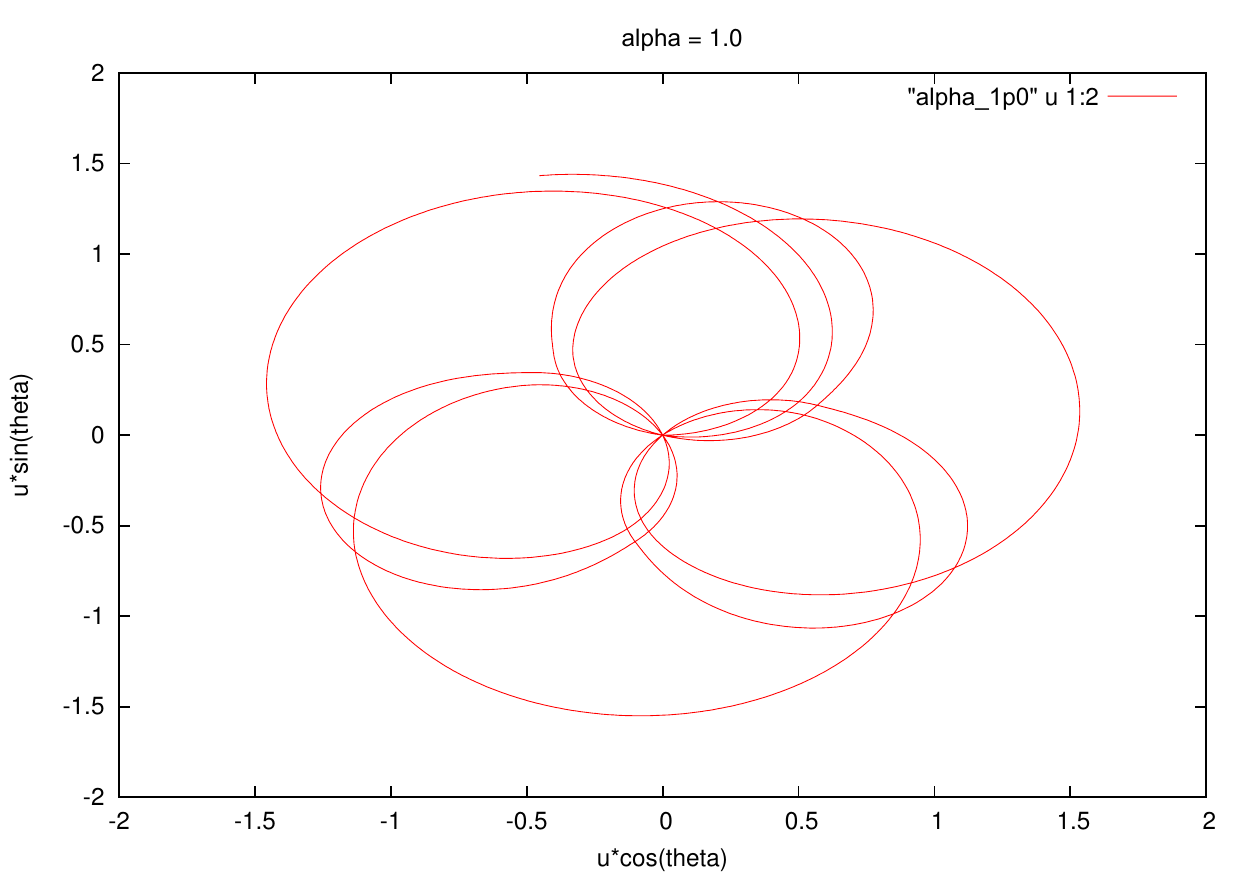}
\includegraphics[scale=0.6]{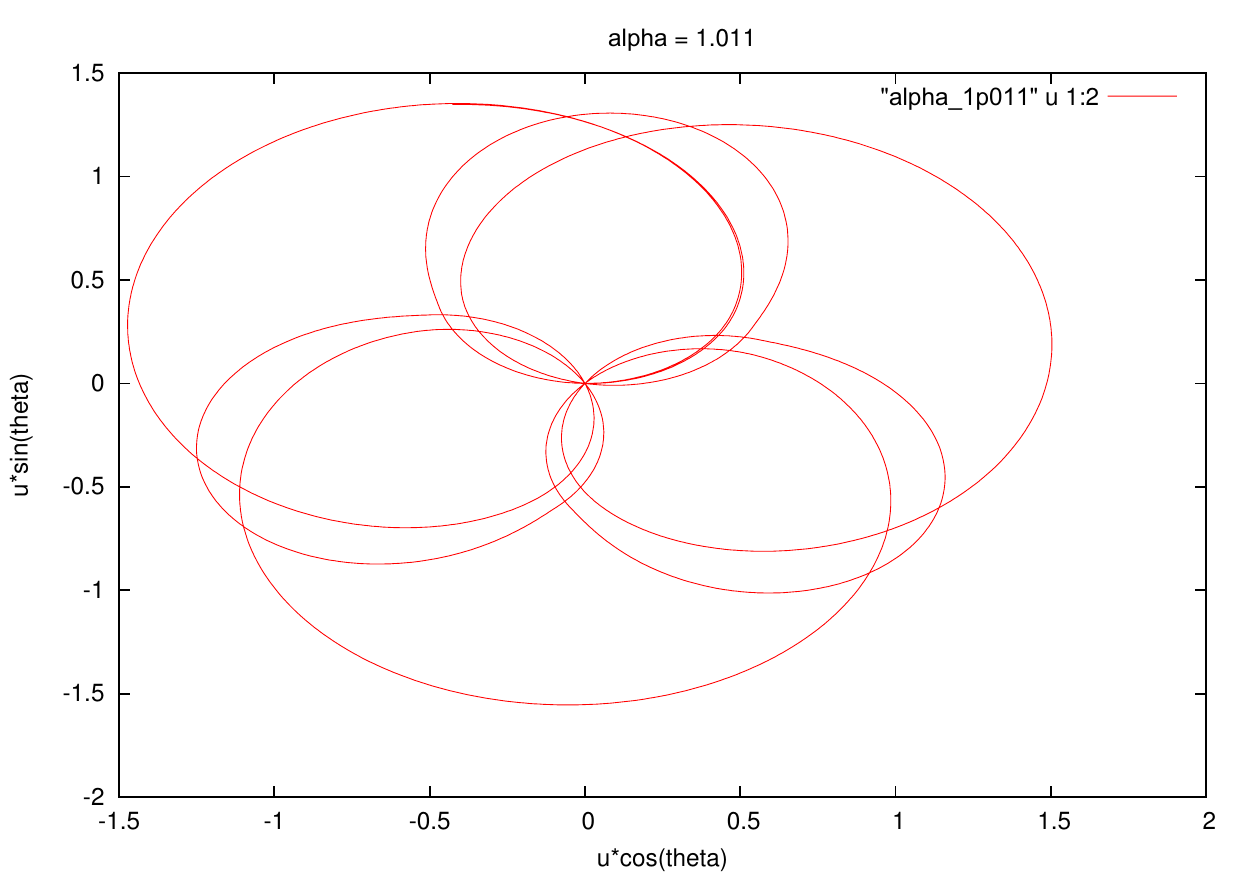}
\caption{Precessing Ellipse with $\alpha=1.0$ and Closed orbits with $\alpha=1.011$}
\end{figure}

For $\alpha$ = 5; we observe two distinct classes of petals, one within the other. Maintaining the two distinct petal structures we have been able to find the closed bound orbits for some parameter value ($\alpha$).

\begin{figure}[H]
\includegraphics[scale=0.6]{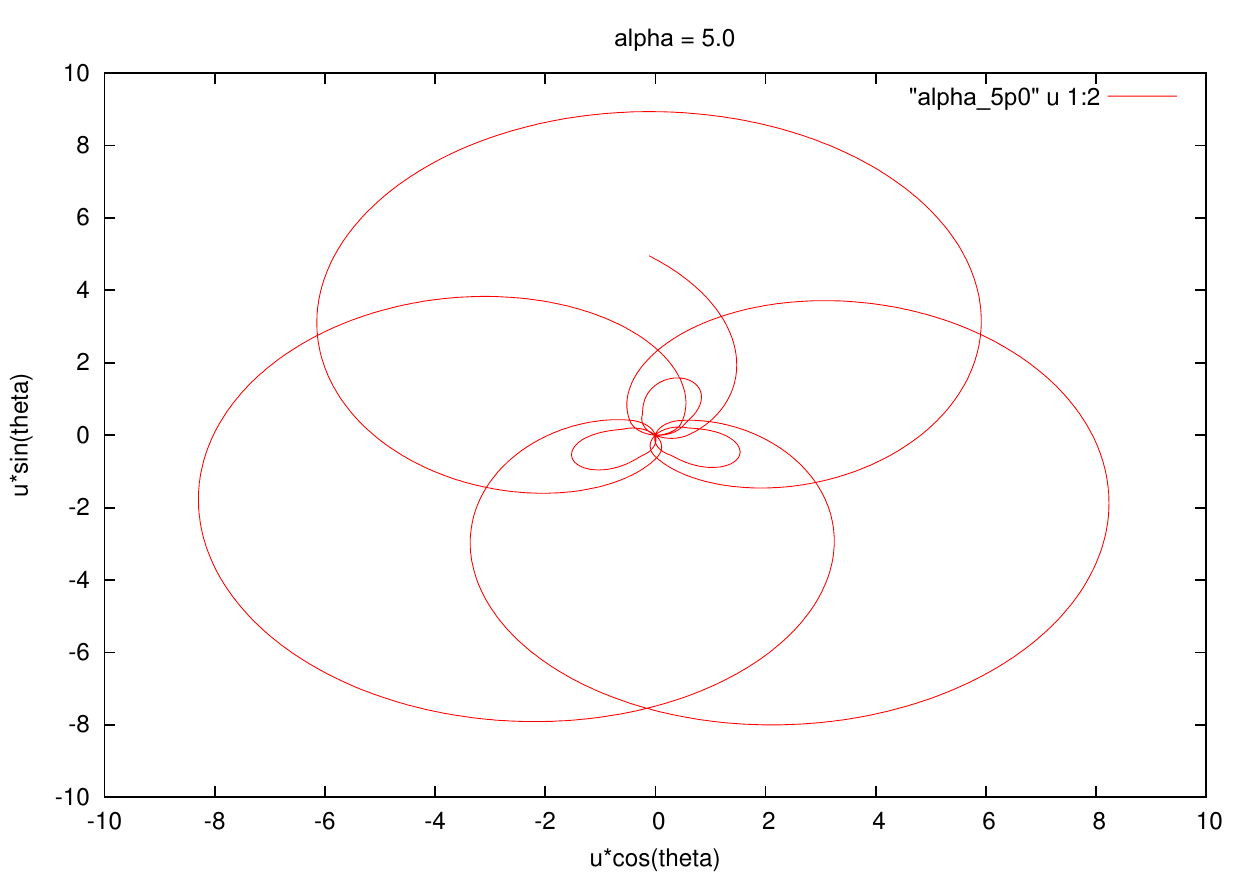}
\includegraphics[scale=0.6]{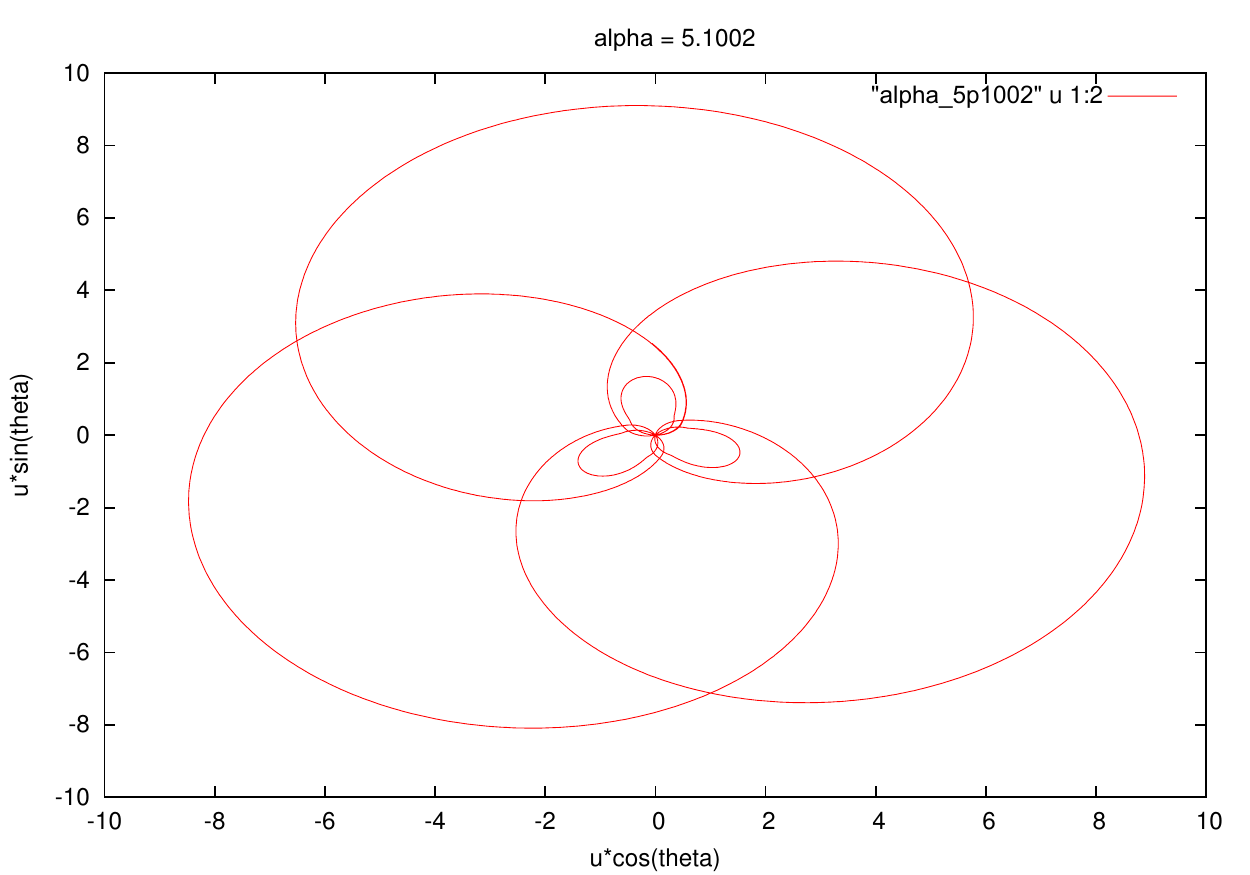}
\caption{Precessing Ellipse with $\alpha=5.0$ and Closed orbits with $\alpha=5.1002$}
\end{figure}

For $\alpha$ = 10 also; we observe two distinct classes of petals but the ratio of the bigger to smaller orbit has changed. Here also we have observed closed orbits.

\begin{figure}[H]
\includegraphics[scale=0.6]{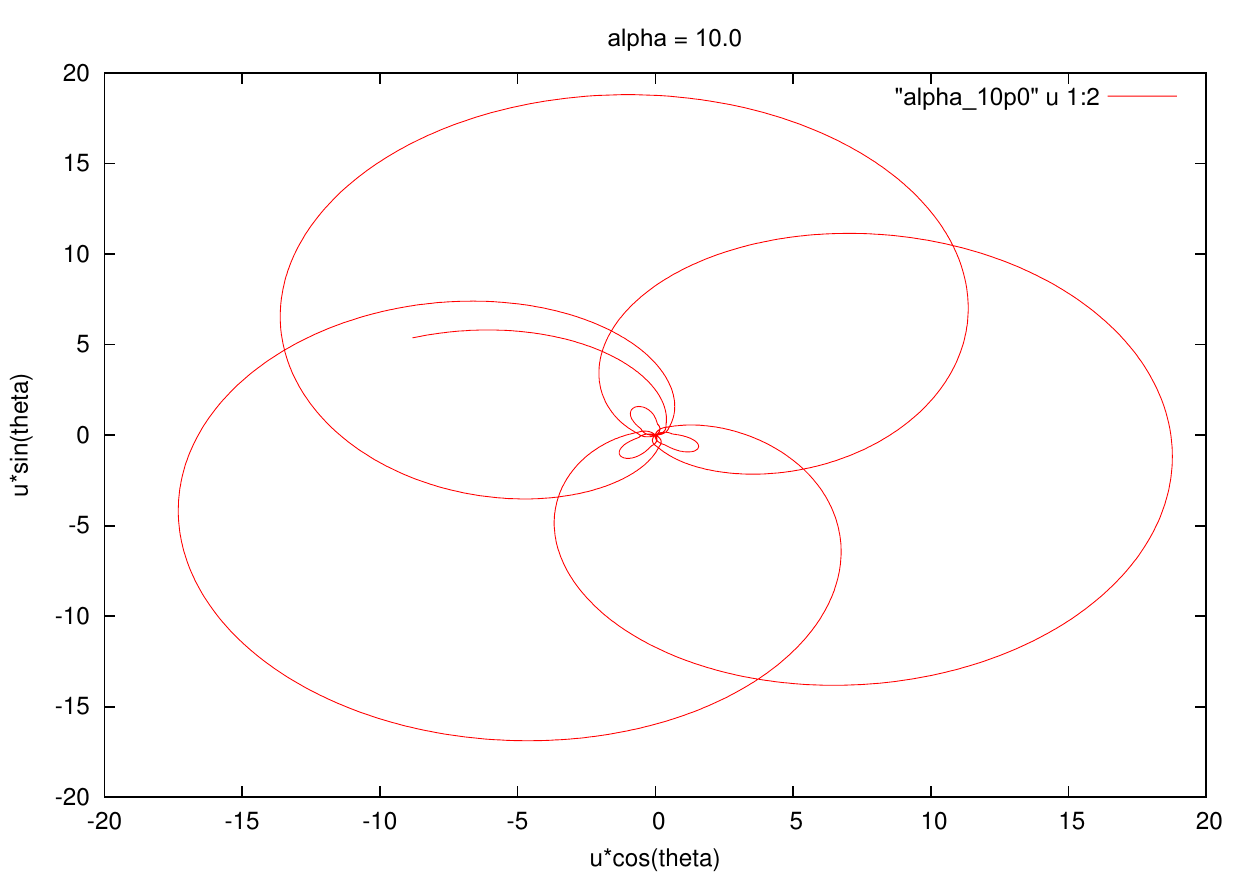}
\includegraphics[scale=0.6]{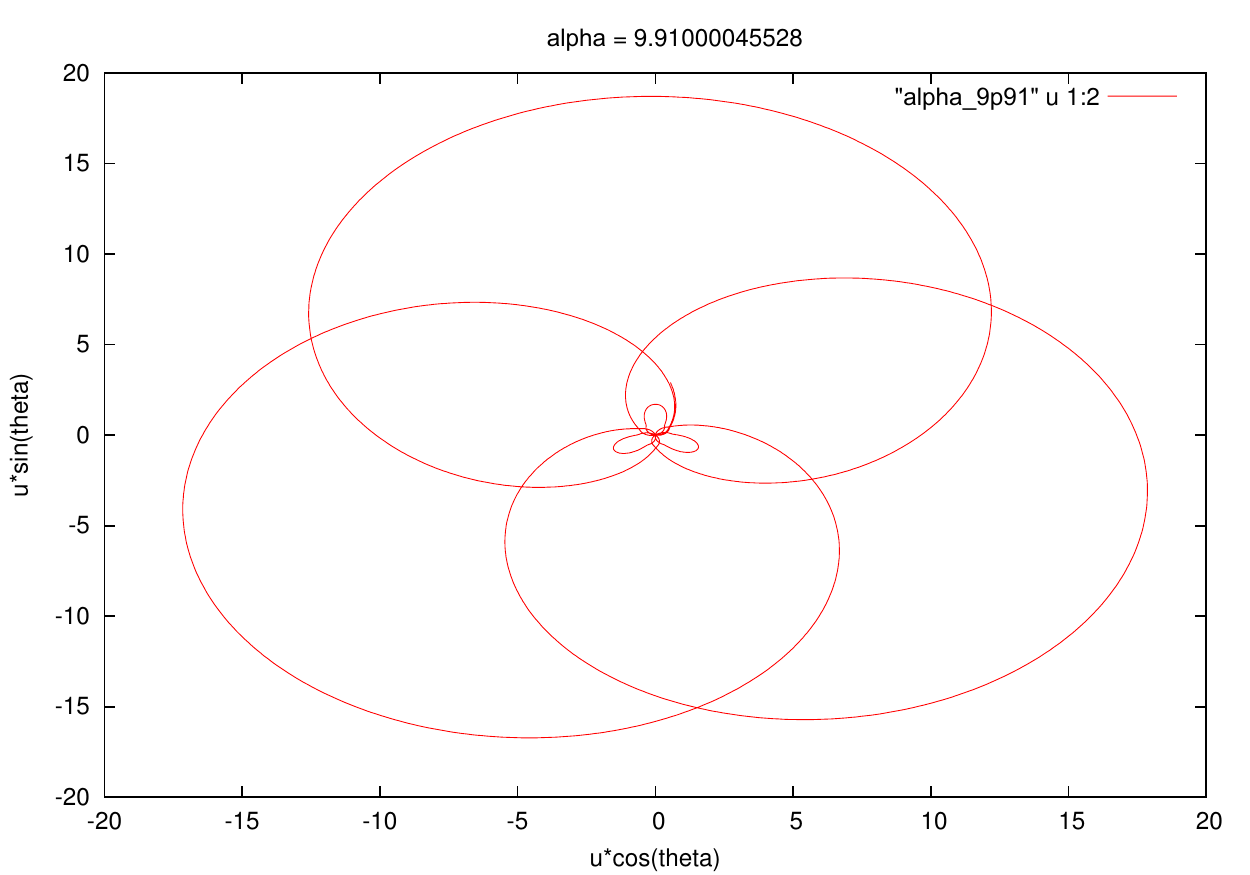}
\caption{Precessing Ellipse with $\alpha=10.0$ and Closed orbits with $\alpha=9.91000045528$}
\end{figure}

Thus we observe that increasing the magnitude of coupling constant increases the precession speed. In some cases it also changes the nature of the orbits. Thus if one can tune the coupling constant in such a way that  after one complete rotation the precessed orbit matches with the first orbit, it traces back the same path and hence forms a closed bound orbit. The same is true for the orbits with two distinct types of petals. Thus for Yukawa potential boundness criteria is just a cautious choice of the coupling parameter that we have presented extensively for many parameter (coupling constant; $\alpha$) values in the above figures.\\

\section{Energy diagrams for yukawa potential:}

Now we calculate the threshold energy for the bound orbits. For a constant value of coupling constant, $\alpha = 0.05$, we plot equation $\ref{en}$ for different values of energy. ($h = 2,1,0.5,0.25,0.1,0.05,0.005$). Note that all the curves cut the negative $u$ axis perpendicularly representing $\frac{dV_{eff}}{du}$ to be undefined. This states that we have not reached the threshold energy value. For energy values less than $0.5$ we observe no crossing in the negative $u$ axis representing that we have gone further below the threshold energy for coupling constant $\alpha = 0.05$

\begin{figure}[H]
\centering
\includegraphics[scale=0.8]{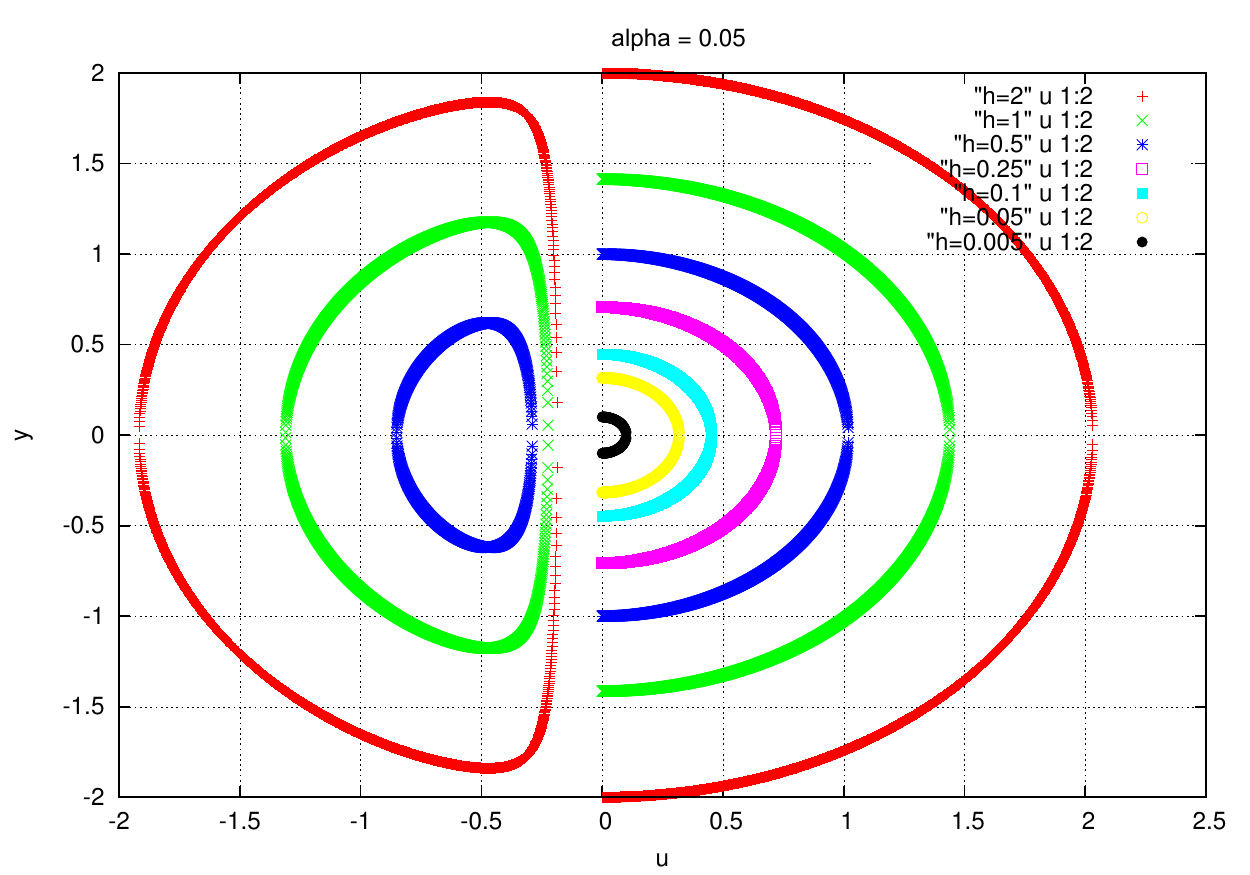}
\caption{Energy diagram for constant coupling parameter $\alpha = 0.05$ and varying energy $h$}
\end{figure}

Again for a constant $h = 1.5$ we plot for different values of coupling constant $(\alpha = 0.0005, 0.005, 0.05, 0.5, 1)$

\begin{figure}[H]
\centering
\includegraphics[scale=0.8]{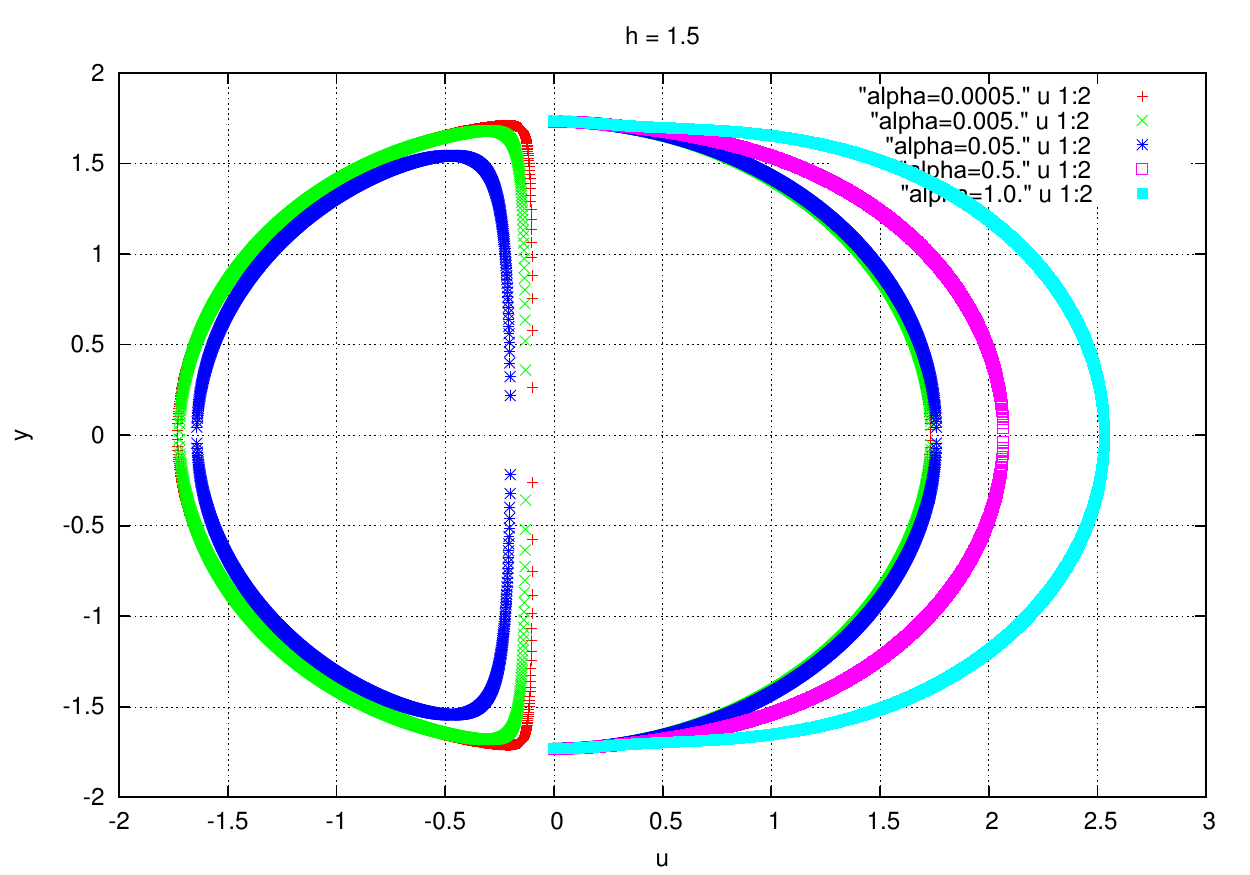}
\caption{Energy diagram with constant energy $h = 1.5$ and varying coupling parameter $\alpha$}
\end{figure}

Now we calculate the minimum values of energy and the u-nullclines graphically. For a constant value of $\alpha$ we start with a bound orbit in the negative $u$ axis and keep on decreasing energy untill the circle reduces to a point. Then if the value of energy is further decreased; suddenly we observe that the point in the negative $u$ axis disappears. This corresponds to the minimum value of energy for a closed orbit in the $u-\theta$ diagram. In the figure below we have plotted the minimum value of energy ($h$) upto which the point in the negative $u$ axis was found. The position of the point gives the value of the $u$-nullcline and we have measured that value for different $\alpha$.

\begin{figure}[H]
\centering
\includegraphics[scale=0.8]{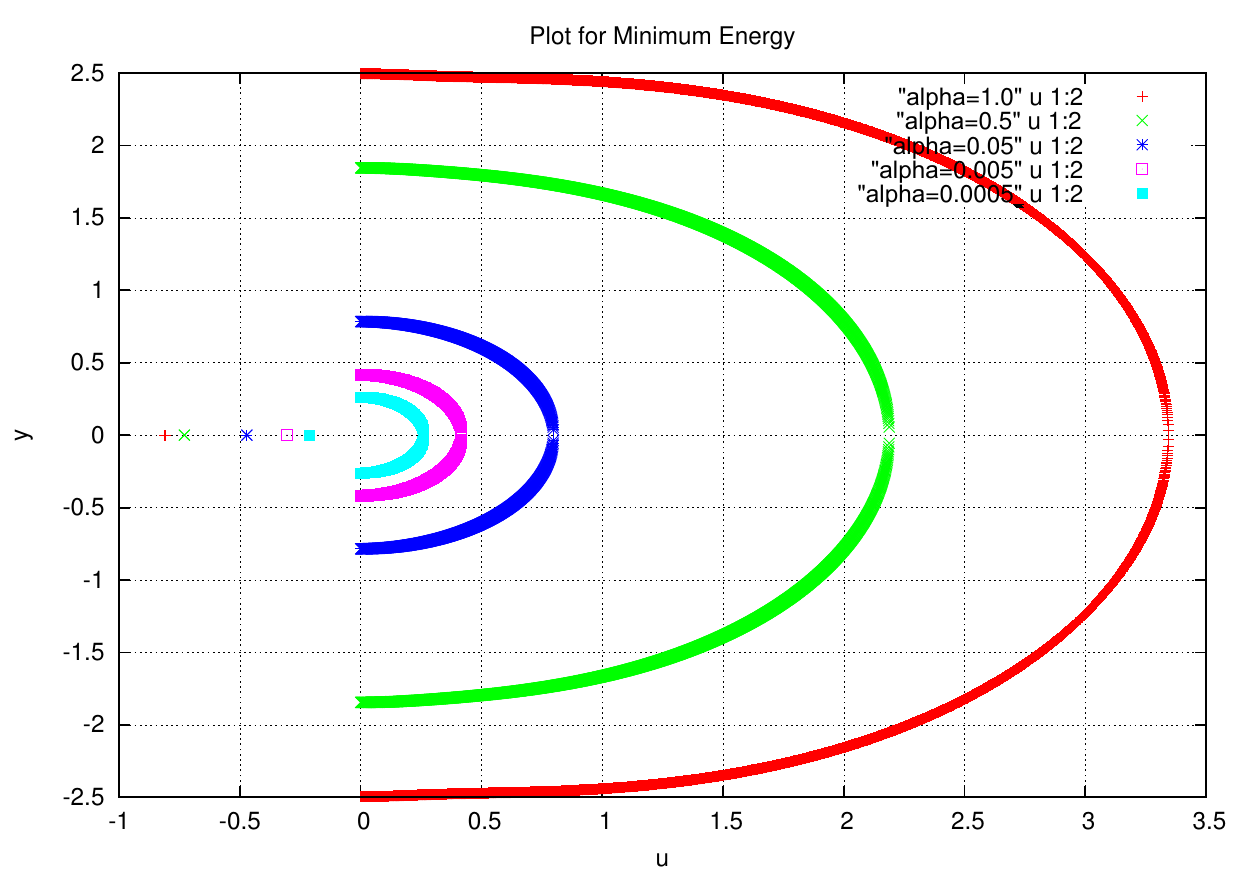}
\caption{Threshold energy values for different coupling parameters $\alpha$}
\end{figure}

From the above graph we have made a table for coupling constant and minimum threshold energy for a bound orbit (that we have obtained numerically by varying the parameter $h$) and the u-nullcline.\\

\begin{center}
\begin{tabular}{||l|p{140pt}|p{100pt}||}
\hline \hline
{\bf Value of $\alpha$} & {\bf Minimum value of $h(+ve)$ for which closed curve occurs in $ - ve$ axis
} & {\bf Value of u}  \\
\hline
0.0005	 & 0.0343254079 & -0.211397\\
\hline
0.005 & 0.086987336 & -0.30454214\\
\hline
0.05 & 0.30773621 & -0.47066493\\
\hline
0.5 & 1.7026556091 & -0.729614\\
\hline 
1.0 & 3.1119334 & -0.80938\\
\hline \hline
\end{tabular}
\newline
\vskip 0.2cm
{ Table 1: Minimum values of energy for bound orbits\\ and the u-nullclines}
\end{center}

Now from the equation $\frac{dV_{eff}}{du} = 0 $ we get:
\begin{equation*}
u - \alpha \left(1 + \frac{1}{u} \right) e^{-\frac{1}{u}} = 0
\end{equation*}
and solving this equation analytically as well as graphically we get the table below. The graphical solution is given for a comparison.

\begin{figure}[H]
\centering
\includegraphics[scale=0.8]{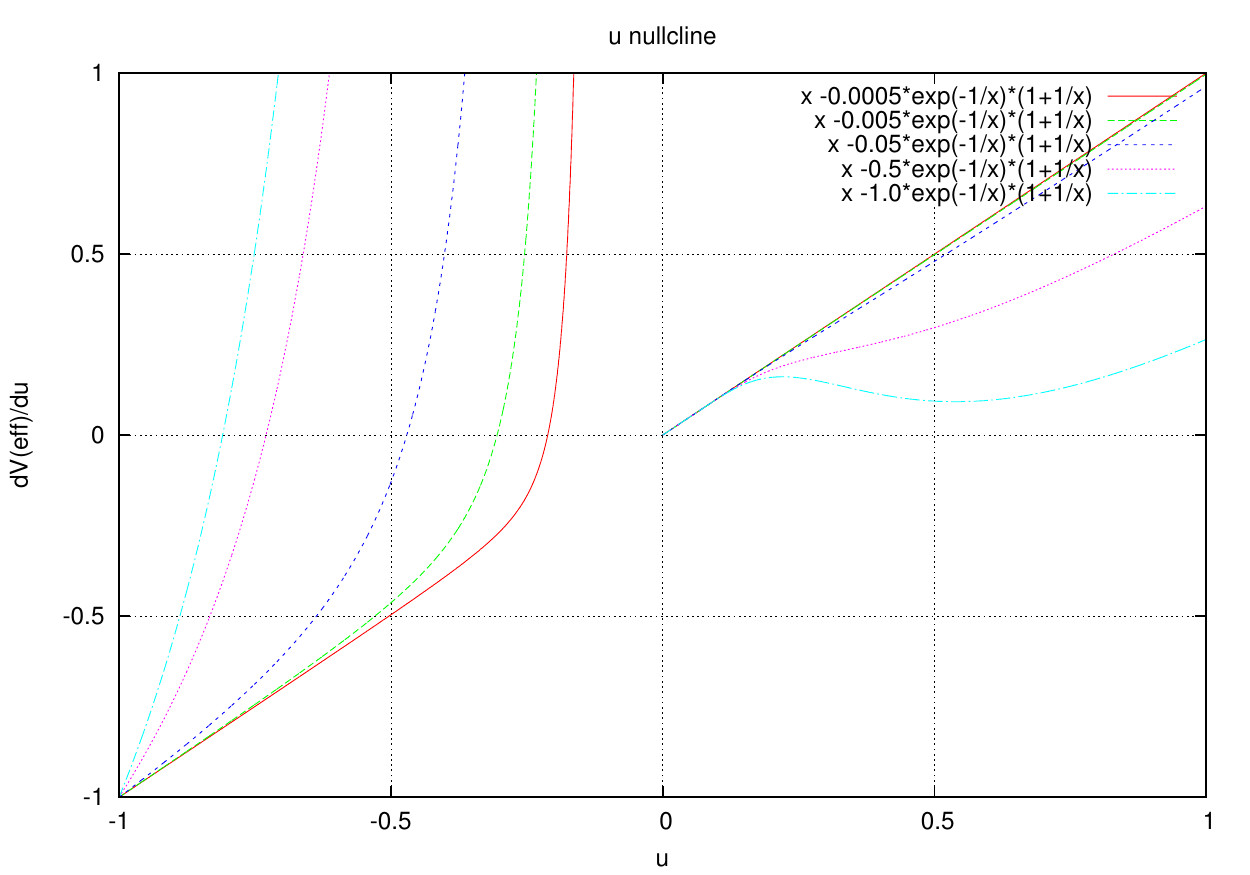}
\caption{Graphical solution of u-nullclines}
\end{figure}

\begin{center}
\begin{tabular}{||l|p{60pt}||}
\hline \hline
{\bf Value of $\alpha$} & {\bf Value of u}  \\
\hline
0.0005 & -0.211397\\
\hline
0.005 & -0.30454214\\
\hline
0.05 & -0.47066493\\
\hline
0.5 & -0.729614\\
\hline 
1.0 & -0.80938\\
\hline \hline
\end{tabular}
\end{center}
\begin{center}
{ Table 2:the u-nullclines \\ vs the coupling constant}
\end{center}
This is exactly in agreement with the value provided above in the Table 1.\\
We also note that $\frac{d^2V_{eff}}{du^2} \textgreater 0$, hence $V_{eff}$ has minima at those points given in the above table for different $\alpha$.\\
Hence the paths shown in the above figures are stable.\\

\section{Conclusion:}
Thus from the above study we have found a set of values of coupling constant for a fixed energy value above a certain threshold limit. So, it is evident that a proper tuning of the coupling constant and energy may change the motion of a particle from closed to aperiodic or vice-versa.\\
A possible application for this phenomena is the removal of outermost electrons of a heavy atom. The outermost electrons experience a screened coulomb or yukawa type of interaction due to the presence of other inner electrons. Thus tuning the strength of external magnetic field, one can tune the angular momentum ($J$)of the outermost electron which in turn affects the coupling constant ($\alpha = \frac{m\alpha}{J^2a}$). For a fixed energy, a proper choice of magnetic field can cause easy removal of outermost electrons from their shells. Very heavy elements or atoms in very excited states, can be exposed to external electric field that will alter the energy of the outermost electron, keeping the angular momentum conserved, so that its orbit becomes aperiodic. One can also think of electrons just below the conduction band in a metal undergoing some periodic orbits, can be taken up to the conduction band by modulating the coupling constant as mentioned above.\\
In case of strongly-coupled complex plasma the caging effect on the dust particles can give rise to some of these closed orbits. This in turn may lead to some oscillating collective behaviors in complex plasmas.\\ 
Another important application of yukawa potential is in astronomy to explain anomaly in the period of orbits of planets$[\ref{anomaly}]$, mean motion of planets in solar system$[\ref{solar}]$ and in long time run of satellites$[\ref{satellite}]$. For astrophysical measurements based on radar signals near the sun, get also affected by the yukawa correction term in gravitational potential$[\ref{radar}]$. For all the above cases the periodicity and closure issues discussed above are quite relevant. 

\vskip 5mm

{\Large \bf {Acknowledgements:}}\\

RM and SS are thankful to Mrityunjay Kundu$^{\ref{ipr}}$, Sayantani Bhattacharyya$^{\ref{rkmvu}}$ (presently at IIT Kanpur) and Abhijit Sen$^{\ref{ipr}}$ for their valuable suggestions and discussions. The authors also thank an anonymous referee of Indian Journal of Physics for several insightfull suggestion.\\

\vskip 5mm

{\Large \bf {References:}}
\begin{enumerate}
\item \label{land} L D Landau and E M Lifshitz {\it {Course of Theoretical Physics (Mechanics)}} (Pergamon Press : Oxford) {\bf{Vol 1}} Ch 3, Sec 14, p 32 (1969)
\item \label{prev1} I Rodriguez and J L Brun {\it { Eur. J. Phys.}} {\bf{19}} 41 (1998).
\item \label{prev2} J L Brun and A F Pacheco {\it {Celestial Mech. Dyn. Astr}} {\bf{96}} 311 (2006)
\item \label{linear} M Malek {\it {Nonlinear Systems of Ordinary Differential Equations}} (California State University, East Bay) 
\item \label{Ross} S L Ross {\it {Differential Equations}} (John Willey \& Sons) Ch 13, Sec 13.3, p 661 (1984)
\item C Sparrow {\it { The Qualitative Theory of Ordinary Differential Equations}} (Lecture Notes : MA371) p 15 (2009)
\item \label{strogatz} S H Strogatz {\it {Nonlinear Dynamics and Chaos}} (Levant Books) Ch 8, Sec 8.2, p 253 and Ch 5, Sec 5.2, p 137
\item \label{anomaly}I Haranas, O Ragos and V Mioc {\it {Astrophys Space Sci}} {\bf{332}} 107 (2011)
\item \label{satellite}I Haranas and O Ragos {\it {Astrophys Space Sci}} {\bf{331}} 115 (2011)
\item \label{analytical} E Fischbach and C L Talmadge {\it {The Search for Non-Newtonian Gravity}} (Springer) p 113 (1999)
\item \label{Harbasic} J W Moffat and V T Toth {\it {ArXiv:0712.1796v5 [gr-qc]}} (2009)
\item \label{appln} L Iorio {\it {Scholarly Research Exchange}} {\bf{2008}} Article ID 238385
\item \label{solar} I Haranas, I Kotsireas, G Gomez, M J Fullana, I Gkigkitzis {\it {Astrophys Space Sci}} {\bf{361:365}} (2016)
\item I Haranas, I Gkigkitzis {\it {Astrophys Space Sci}} {\bf{337:693-702}} (2012)
\item L Iorio and M L Ruggiero {\it {Int. J. Mod. Phys. A}} {\bf{22}} (No. 29) 5379-5389 (2007)
\item \label{radar} I Haranas, O Ragos {\it {Astrophys Space Sci}} {\bf{334:71-74}} (2011)
\item O Bertolami {\it {Int. J. Mod. Phys. A}} {\bf{16}} (No. 12A) 2003-2012 (2007)
\end{enumerate}

\end{document}